\documentclass[review]{elsarticle}

\usepackage[ruled,vlined]{algorithm2e}

\usepackage{rotating}
\usepackage{lineno,hyperref}
\modulolinenumbers[5]

\usepackage{amsfonts}
\usepackage{graphicx}
\usepackage{epstopdf}
\usepackage{amsopn}
\usepackage{natbib}

\usepackage{hyperref}
\usepackage{amsmath}

\usepackage{geometry} 
\usepackage{fleqn}
\usepackage{graphicx} 
\usepackage{endfloat}

\usepackage{algorithmic}

\ifpdf
  \DeclareGraphicsExtensions{.eps,.pdf,.png,.jpg}
\else
  \DeclareGraphicsExtensions{.eps}
\fi
\usepackage{mathtools}

\usepackage{tikz}
\usetikzlibrary{quantikz}

\journal{Journal of \LaTeX\ Templates}

\begin{document}

\begin{frontmatter}

\title{Special Functions and HHL Quantum Algorithm for Solving Moving Boundary Value Problems Occurring in Electric Contact Phenomena 
}


\author[mymainaddress,mysecondaryaddress,mythirdaddress]{Merey M. Sarsengeldin\corref{mycorrespondingauthor}}
\cortext[mycorrespondingauthor]{Corresponding author}
\ead{merey.sarsengeldin@ucf.edu, dr.sarsengeldin@gmail.com}


\author[mymainaddress]{Zuhair M. Nashed}

\address[mymainaddress]{Department of Mathematics, University of Central Florida, Orlando, FL, US}
\address[mysecondaryaddress]{National Academy of Sciences, Institute of Mathematics and Mathematical Modeling, Almaty, Kazakhstan}
\address[mythirdaddress]{Satbayev University, Almaty, Kazakhstan}

\begin{abstract}
This is a series of studies devoted to modeling and solving heat and mass transfer problems occurring in electric contacts where we employ and develop mathematical apparatus along with quantum algorithms for solving moving boundary value problems. In this particular study we utilize special functions and demonstrate the use of Harrow-Hassidim-Lloyd (HHL) quantum algorithm for finding exact and approximate solutions of Generalized Heat Equation with moving boundaries and as examples we consider plane and spherical cases. In spherical case the Generalized Heat Equation is reduced to linear moving boundary value problem with discontinuous coefficients and solved exactly. In plane case we use collocation method for approximate solution of Inverse Two-Phase Stefan problem.
\end{abstract}


\end{frontmatter}

\linenumbers

\section{Introduction}
Electrical contacts, their design and reliability play crucial role in designing modern electrical apparatuses. A lot of electric contact phenomena accompanied with heat and mass transfer like arcing and bridging are very rapid (nanosecond range)\cite{holm2013electric, slade2017electrical} that their experimental study is very difficult or sometimes impossible and the need of their mathematical modeling is due not only to the need to optimize the planning experiment, but also sometimes due to the impossibility to use a different approach. Free (FBVPs) and Moving Boundary Value Problems (MBVPs) take in account phase transformations  \cite{friedman1970,rubinstein2000stefan}, agree with experimental data and can serve as models for aforementioned processes \cite{khar15holm,khar16ICEC,kharin2012}.\\
From theoretical point of view, these problems are among the most challenging problems in the theory of non-linear parabolic equations, which along with the desired solution an unknown moving boundary has to be found. In some specific cases it is possible to construct Heat Potentials for which, boundary value problems can be reduced to integral equations \cite{friedman1970,rubinstein2000stefan,tikhonov2013equations}. However, in the case of domains that degenerate at the initial time, there are additional difficulties due to the singularity of integral equations, which belong to the class of pseudo - Volterra equations which can be solved in special cases and hard to solve in general case. A reader can refer to the long list of studies in \cite{Tarzia2000ABO} and literature therein dedicated to the MBVPs. Despite the great value and exhaustiveness of all these results, investigation and elaboration of both exact and approximate methods for solving MBVPs responsible for adequate modeling electric contact phenomena  is still an actual mathematical problem.\\
In this paper we consider a class of PDEs with moving boundaries
\begin{align}
&\frac{\partial {\theta}}{\partial {t}} = a^2\left(\frac{\partial {\theta^2}}{\partial {x^2}} + \frac{\nu}{x}\frac{\partial{\theta}}{\partial{x}} \right),\hspace{1em} \alpha(t)<x<\beta(t),\hspace{1em}  -\infty<\nu<\infty,\hspace{1em} t>0, \label{geneq}\
\end{align}
which can be solved by the series of linear combinations of special functions which apriori satisfy equation \ref{geneq}
\begin{align}
&S_{\gamma,\nu}^{1}\left(x,t\right) = \left(2a\sqrt{t}\right)^{\gamma}\Phi\left(-\frac{\gamma}{2},\frac{\nu+1}{2};-\frac{x^{2}}{4a^{2}t}\right),\hspace{1em} -\infty<\gamma,\nu<\infty,\hspace{1em} \label{hyper1}\\[1mm]
&S_{\gamma,\nu}^{2}\left(x,t\right) = \left(2a\sqrt{t}\right)^{\gamma}\left(\frac{x^{2}}{4a^{2}t}\right)^{\frac{1-\nu}{2}}\Phi\left(\frac{1-\nu-\gamma}{2},\frac{3-\nu}{2};-\frac{x^{2}}{4a^{2}t}\right).\label{hyper2}\hspace{1em}\
\end{align}
Generalized Heat Equation and its solutions were studied in \cite{appell,widder75,haimo68,widapptransf63,brag65}, and was successfully applied in \cite{khar15holm,khar16ICEC,kharin2012, kharnovosib17} for modeling and solving Heat and Mass transfer problems in diverse electric contact phenomena. Our goal in this series of studies is to develop new computational methods for solving MBVPs where we will be employing and developing quantum algorithms as well.\\
Pioneering studies \cite{ben80,fe82} in 1980s gave a birth to a new paradigm in computation which we call nowadays quantum computing, whereby information is encoded in a quantum system. Further on, in 1990s a series of studies \cite{sho97,gro96,gro97} dedicated to quantum algorithms provided exponential speed-up in run time over the best known classical algorithms for same tasks. In last decades, consistent advances in theory and experiments generated a plethora of powerful quantum algorithms \cite{mo16} 
which surpass their classical counterparts in terms of computational power, however worth noting that their applications are restricted to few use cases due to the challenges related to their physical realization. Careful physical realization may lead to profound results in reaching exponential speed-up.\\
In this particular study, we will be using one of such powerful quantum algorithms developed by Harrow-Hassidim-Lloyd (HHL) \cite{hhl09} to solve MBVPs. The HHL algorithm, its modifications and improvements \cite{hhl09, wosnig18physrevlet,duan20physleta,abhijith18arxiv,chakraborty2018power,Dervovic2018QuantumLS} (both for sparse and dense matrices) is the operator inversion or linear systems solving quantum algorithm, has wide range of applications \cite{duan20physleta} as well as attempts to dequantize them \cite{tang18} 
and provides exponential speed-up over the classical algorithms. For detailed survey on improvements and limitations, complexity, QRAM and physical implementation  of the algorithm we refer reader to \cite{duan20physleta, Dervovic2018QuantumLS} and literature therein.\\
We consider a linear operator equation
\begin{align}
Mx = b,
\label{mateq}
\end{align}
where in this study we assume that $M$ is Hermitian and s-sparse matrix, and b is a vector column. This condition can be relaxed and it can be shown that $\tilde{M} = \begin{bmatrix}
0 & M\\
M^{T} & 0
\end{bmatrix}$ can be brought to Hermitian matrix.
Since $\tilde{M}$ is Hermitian, we can solve the equation 
$\tilde{M}\vec{y} = \begin{bmatrix}
\vec{b}\\
0
\end{bmatrix}$
to obtain $y = \begin{bmatrix}
0\\
\vec{x}
\end{bmatrix}$.
Therefore the rest of the article we assume that $M$ is Hermitian.\\
The idea of the method is to reduce given MBVP to the equation \ref{mateq} and apply HHL algorithm.
In this study we will consider an "ideal" case where the data is encoded "efficiently" and refer reader to \cite{Dervovic2018QuantumLS} and literature therein for different methods of Hamiltonian simulation and quantum phase estimation.
\section{Main results}
\label{sec:main}
Equation \ref{geneq} with arbitrary $\nu$ is a generalized heat equation which can serve as a model for bridging processes in electrical contacts with variable cross section. For $\nu=0,1,2$ equation \ref{geneq} is transformed to the following heat equation in linear, spherical and cylindrical coordinates respectively
\begin{align}
&\frac{\partial {\theta}}{\partial {t}} = a^2\frac{\partial {\theta^2}}{\partial {x^2}},\hspace{1em} &\alpha(t)<x<\beta(t),\hspace{1em} t>0 \label{lin}\\[1mm]
&\frac{\partial {\theta}}{\partial {t}} = a^2\left(\frac{\partial {\theta^2}}{\partial {x^2}} + \frac{1}{x}\frac{\partial{\theta}}{\partial{x}} \right),\hspace{1em}  &\alpha(t)<x<\beta(t),\hspace{1em} t>0 \label{cyl}\\[1mm]
&\frac{\partial {\theta}}{\partial {t}} = a^2\left(\frac{\partial {\theta^2}}{\partial {x^2}} + \frac{2}{x}\frac{\partial{\theta}}{\partial{x}} \right),\hspace{1em}  &\alpha(t)<x<\beta(t),\hspace{1em} t>0 \label{sphr}
\end{align}
and from \ref{hyper1} and \ref{hyper2} one can obtain solutions for equations \ref{lin},\ref{cyl} and \ref{sphr} in the form of following series of linear combinations of special functions 
\begin{align}
&\theta(x,t)= \sum_{n=0}^{k}\left(2a\sqrt{t}\right)^{n}\left[A_ni^nerfc\left( \frac{x}{2a\sqrt{t}} \right)+ B_ni^nerfc\left( \frac{-x}{2a\sqrt{t}} \right)\right] \label{sollin}\\[3mm]
&\theta(x,t)= \sum_{n=0}^{k}F_n\frac{n!}{\left(2n\right)!}\left(4a^2t\right)^{n}L_n\left(-\frac{x}{4a^{2}t}\right), \label{solcyl}\\[3mm]
&\theta(x,t)= \frac{1}{x}\sum_{n=0}^{k}\left(2a\sqrt{t}\right)^{n}\left[C_ni^nerfc\left( \frac{x}{2a\sqrt{t}} \right)+ D_ni^nerfc\left( \frac{-x}{2a\sqrt{t}} \right)\right], \label{solsphr}
\end{align}
where coefficients $A_n,B_n,C_n,D_n,F_n$ and $k$ have to be determined and can be found from boundary and initial conditions subject to corresponding equations \ref{lin},\ref{cyl} and \ref{sphr} by using quantum HHL algorithm. After substituting solution functions into boundary conditions, the problem is reduced to the system of linear algebraic equations which are solved by the HHL algorithm. As for arbitrary moving boundary $\alpha(t)$ coefficients of solution functions are calculated in the same manner in combination with Faa Di Bruno's formula and HHL  Algorithm.
Following formula is useful for determining coefficients in \ref{sollin}, \ref{solcyl} and \ref{solsphr} from initial conditions of corresponding MBVPs considered in following sections
\begin{align}
\underset{x\to 0}{\lim}\frac{1}{z^{\beta}}\Phi\left(-\frac{\beta}{2},\mu;-z^{2}\right) = \frac{\Gamma(\mu)}{\Gamma(\mu+\frac{\beta}{2})}.\label{crlry}\
\end{align}
Exact or approximate solutions of \ref{geneq} for arbitrary $\nu$ can be reduced to system of linear algebraic equations or \ref{mateq}. The HHL algorithm applied for solving MBVPs provides exponential speedup in run time over the classical algorithm under the ideal case assumption i.e. Hamiltonian simulation, phase estimation, load and read out of data are implemented "efficiently".\\
\begin{algorithm}[H]
\SetAlgoLined
\KwData{Load the data $\ket{b}\in\mathbb{C} ^{N}$}
\KwResult{Apply an observable  M  to calculate $F(x)=\bra{x}M\ket{x}.$}
initialization\;
\While{outcome is not $1$ }{
\begin{itemize} 
  \item Apply Quantum Phase Estimation (QPE) with\\ $U=e^{iMt}:=\sum_{j=0}^{N-1}e^{i\lambda_{j}t}\ket{u_j}\bra{u_j}$. Which implies $\sum_{j=0}^{N-1}b_{j}\ket{\lambda_j}_{n_{l}}\bra{u_j}_{n_{b}}$, \\ in the eigenbasis of  $M$\\
where $\ket{\lambda_j}_{n_{l}}$ is the  $n_{l}$ -bit binary representation of $\lambda_{j}$ .
  \item Add an ancilla qubit and apply a rotation conditioned on  $\ket{\lambda_j},$\\
  $\sum_{j=0}^{N-1}b_{j}\ket{\lambda_j}_{n_{l}}\bra{u_j}_{n_{b}}\left(\sqrt{1-\frac{C^2}{\lambda_j^2}}\ket{0}+\frac{C}{\lambda_j}\ket{1}\right)$, $C$ - normalization constant.
  \item Apply $QPE^{\dag}.$ This results in\\
  $\sum_{j=0}^{N-1}b_{j}\ket{0}_{n_{l}}\bra{u_j}_{n_{b}}\left(\sqrt{1-\frac{C^2}{\lambda_j^2}}\ket{0}+\frac{C}{\lambda_j}\ket{1}\right);$
\end{itemize}
\eIf{If the outcome is  $1$ , the register is in the post-measurement state\\
    $\left(\sqrt{\frac{1}{\sum_{j=1}^{N-1}\left|b_j\right|^2/\left|\lambda_j\right|^2/}}\right)\sum_{j=0}^{N-1}\frac{b_{j}}{\lambda_{j}}\ket{0}_{n_{l}}\bra{u_j}_{n_{b}}$ }{Apply an observable  M  to calculate $F(x)=\bra{x}M\ket{x}$;
}{
repeat the loop\;
}
}
\caption{Quantum HHL Algorithm in Qiskit}
\label{alg:genhhl}
\end{algorithm}


For computational purposes we use Qiskit and IBM Q. Let's consider exact and approximate solutions of two model problems where we demonstrate the use of HHL quantum algorithm. The error of the approximate solution can be estimated by the Maximum Principle.\\


\subsection{HHL algorithm for exact solution of system of MBVP with discontinuous coefficients}
\label{sec:exact}
For electric contacts with small contact surface area (with contact radius $b < 10^{-4} \hspace{1mm} m.$) and low electric current, Holm’s ideal sphere \cite{holm2013electric} and following system of spherical heat equations ($\nu=2$ in \ref{geneq}) can be sufficient for adequate modeling and investigation of diverse electric contact phenomena for example heat transfer in closed electric contacts where $\theta_{1}$ and $\theta_{2}$ are temperature functions in liquid and solid phases respectively.
\begin{align}
&\frac{\partial {\theta_{1}}}{\partial {t}} = a_{1}^{2}\left(\frac{\partial {\theta_{1}^{2}}}{\partial {x^{2}}} + \frac{2}{x}\frac{\partial{\theta_{1}}}{\partial{x}} \right),\hspace{1em}  &b<x<\alpha(t),\hspace{1em} t>0 \label{mod1_eq1}\\[1mm]
&\frac{\partial {\theta_{2}}}{\partial {t}} = a_{2}^{2}\left(\frac{\partial {\theta_{2}^{2}}}{\partial {x^2}} + \frac{2}{x}\frac{\partial{\theta_{2}}}{\partial{x}} \right),\hspace{1em}  &\alpha(t)<x<\infty,\hspace{1em} t>0 \label{mod1_eq2}
\end{align}
where $b$ is the radius of the ideal Holm's sphere.
Using substitution $\theta = \frac{U}{x}+T_{m}$ in equations \ref{mod1_eq1} and \ref{mod1_eq2}  
subject to certain boundary conditions depending on the studied phenomenon can be reduced to the problem below. $T_{m}$ is the melting temperature at moving boundary. For the sake of simplicity we omit Stefan condition and consider $\alpha\sqrt{t}$ moving boundary function which is a good approximation and widely used in applied problems. Let's consider following abstract MBVP:
\begin{align}
&\frac{\partial U_1}{\partial t} = a_1^2\frac{\partial U_1^2}{\partial x^2},\hspace{1em} &0<x<\alpha\sqrt{t},\hspace{1em} t>0,  \label{mod1_eq11}\\[3mm]
&\frac{\partial U_2}{\partial t} = a_2^2\frac{\partial U_2^2}{\partial x^2},\hspace{1em} &\alpha\sqrt{t}<x<\infty,\hspace{1em} t>0, \label{mod1_eq22}\\[3mm]
&U_1(0,0) = 0,\label{mod1_incond1}\\[3mm]
&U_2(x,0) = f(x),\label{mod1_incond2}\\[3mm]
&\left.\frac{\partial U_1}{\partial x} \right|_{x=0} = P(t),  \label{mod1_flux}\\[3mm]
&\left.\sigma\frac{\partial U_1}{\partial x} \right|_{x=\alpha\sqrt{t}} = \left.\frac{\partial U_1}{\partial x} \right|_{x=\alpha\sqrt{t}},  \label{mod1_flux2}\\[3mm]
&U_1(\alpha\sqrt{t},t) = U_2(\alpha\sqrt{t},t)\label{mod1_temp}\\[3mm]
&U_2(\infty,0) = 0.\label{infty}
\end{align}
We represent solution in the form of series
\begin{align}
&U_1(x,t)= \sum_{n=0}^{k}\left(2a_{1}\sqrt{t}\right)^{n}\left[A_ni^nerfc\left( \frac{x}{2a_{1}\sqrt{t}} \right)+ B_ni^nerfc\left( \frac{-x}{2a_{1}\sqrt{t}} \right)\right], \label{mod1_sol1}\\[3mm]
&U_2(x,t)= \sum_{n=0}^{k}\left(2a_{2}\sqrt{t}\right)^{n}\left[C_ni^nerfc\left( \frac{x}{2a_{2}\sqrt{t}} \right)+ D_ni^nerfc\left( \frac{-x}{2a_{2}\sqrt{t}} \right)\right], \label{mod1_sol2}
\end{align}
where $k$ is defined from boundary conditions. To find $D_{n}$ we substitute \ref{mod1_sol2} into initial condition \ref{mod1_incond2}. Taking into account that $\underset{t\to 0}{\lim}\sum_{n=0}^{k}\left(\frac{x}{2a\sqrt{t}}\right)C_{n}i^nerfc\left(\frac{x}{2a\sqrt{t}}\right)=0$, using L'Hopital's rule in $\underset{x\to \infty}{\lim}\frac{i^nerfc(-x)}{x^n}=\frac{2}{n!}$ and in $\underset{t\to 0}{\lim}\frac{\left(2a\sqrt{t}\right)^{n}D_{n}i^nerfc\left(\frac{-x}{2a\sqrt{t}}\right)}{\left(\frac{x}{2a\sqrt{t}}\right)^{n}}\left(\frac{x}{2a\sqrt{t}}\right)^{n}=\frac{2x^{n}D_n}{n!}$, and using formula \ref{crlry} we obtain following expression for $D_{n}$ coefficients
\begin{align}
&\sum_{n=0}^{k}\frac{2x^{n}D_n}{n!}=\sum_{n=0}^{m}f^{n}(0)\frac{x^{n}}{n!}
\label{dn_coeff}
\end{align}
Finally, after comparing coefficients at $x^{n}$ in \ref{dn_coeff} we get following expression for $D_{n}$
\begin{align}
&D_n=\frac{f^{n}(0)}{2}
\label{mod1_Dn}
\end{align}
From conditions \ref{mod1_flux}, \ref{mod1_flux2}, \ref{mod1_temp} and \ref{mod1_sol1}, \ref{mod1_sol2} we get following expressions
\begin{subequations}
\begin{align}
&\left.\frac{\partial U_1}{\partial x} \right|_{x=0} = \sum_{n=0}^{k}\left(2a_{1}\sqrt{t}\right)^{n-1}i^{n-1}erfc\left( 0 \right)\left[-A_{n}+ B_{n} \right]\equiv \sum_{n=0}^{k}P^{n}(0)\frac{t^{\frac{n}{2} }}{n!}, \label{mod1_sys1}\\
&\sigma \sum_{n=0}^{k}\left(2a_{1}\sqrt{t}\right)^{n-1}\left[-A_ni^{n-1}erfc\left( \frac{\alpha}{2a_{1}} \right)+ B_ni^{n-1}erfc\left( \frac{-\alpha}{2a_{1}} \right)\right]\label{mod1_sys2}\\
&=\sum_{n=0}^{k}\left(2a_{2}\sqrt{t}\right)^{n-1}\left[-C_ni^{n-1}erfc\left( \frac{\alpha}{2a_{2}} \right)+ D_ni^{n-1}erfc\left( \frac{-\alpha}{2a_{2}} \right)\right]\nonumber,\\ 
&\sum_{n=0}^{k}\left(2a_{1}\sqrt{t}\right)^{n}\left[A_ni^nerfc\left( \frac{\alpha}{2a_{1}} \right)+ B_ni^nerfc\left( \frac{-\alpha}{2a_{1}} \right)\right]\label{mod1_sys3}\\&=
\sum_{n=0}^{k}\left(2a_{2}\sqrt{t}\right)^{n}\left[C_ni^nerfc\left( \frac{\alpha}{2a_{2}} \right)+ D_ni^nerfc\left( \frac{-\alpha}{2a_{2}} \right)\right]\nonumber. 
\end{align}
\end{subequations}
After transferring expressions $-\sum_{n=0}^{k}\left(2a_{2}\sqrt{t}\right)^{n}C_ni^{n-1}erfc\left( \frac{\alpha}{2a_{2}} \right)$,\\
$\sum_{n=0}^{k}\left(2a_{2}\sqrt{t}\right)^{n}C_ni^{n}erfc\left( \frac{\alpha}{2a_{2}} \right)$ to the left side of \ref{mod1_sys2}, \ref{mod1_sys3} and comparing coefficients at same powers of $\sqrt{t}$, equations \ref{mod1_sys1}, \ref{mod1_sys2} and \ref{mod1_sys3} can be represented in the form of system of linear algebraic equations or in the form of matrix equation \ref{mateq}
where
\begin{align}
&M=\begin{pmatrix}m_{11}&m_{12}&0&0&\dots&0\\m_{21}&m_{22}&m_{23}&0&\dots&0\\m_{31}&m_{32}&m_{33}&0&\dots&0\\\vdots&&&\ddots&&\vdots&\\0&\dots&0&m_{3k+13k+1}&m_{3k+13k+2}&0\\0&\dots&0&m_{3k+23k+1}&m_{3k+23k+2}&m_{3k+23(k+1)}\\0&\dots&0&m_{3(k+1)3k+1}&m_{3(k+1)3k+2}&m_{3(k+1) 3(k+1)}\end{pmatrix},\label{mentries}
\\&x=\begin{pmatrix}A_0\\B_0\\C_0\\\vdots\\A_k\\B_k\\C_k\end{pmatrix}, 
b=\begin{pmatrix} 0\\
\frac{f^{0}(0)}{2}i^{-1}erfc\left( \frac{-\alpha}{2a_{2}} \right)\\
\frac{f^{0}(0)}{2}i^{0}erfc\left( \frac{-\alpha}{2a_{2}} \right)\\\vdots\\\frac{P^{k-1}(0)}{(k-1)!\left(2a_1\right)^{k-1}}\\
\frac{f^{k}(0)}{2a_2^{k-1}}i^{k-1}erfc\left( \frac{-\alpha}{2a_{2}} \right)\\
\frac{f^{k}(0)}{2a_2^{k-1}}i^{k}erfc\left( \frac{-\alpha}{2a_{2}} \right)\end{pmatrix}\nonumber,
\end{align}
where
\begin{align*}
&m_{3k+13k+1}=-i^{k-1}erfc\left( 0 \right), 
m_{3k+13k+2}=i^{k-1}erfc\left( 0 \right),
\\&m_{3k+23k+1}=-\sigma a_{1}^{k-1}i^{k-1}erfc\left( \frac{\alpha}{2a_{1}} \right),
m_{3k+23k+2}= \sigma a_{1}^{k-1} i^{k-1}erfc\left( \frac{-\alpha}{2a_{1}} \right),\\&m_{3k+23k+3}= a_{2}^{k-1} i^{k-1}erfc\left( \frac{\alpha}{2a_{2}} \right),\\&m_{3k+33k+1}=a_1^k i^{k}erfc\left( \frac{\alpha}{2a_{1}} \right),
m_{3k+33k+2}= a_1^k i^{k}erfc\left( \frac{-\alpha}{2a_{1}} \right),\\
&m_{3k+33k+3}= -a_2^k i^{k}erfc\left( \frac{\alpha}{2a_{1}} \right)
\end{align*}
and expressions $i^{n}erfc\left( \frac{\pm \alpha}{2a_{j}} \right)$, $i^{n}erfc\left( 0 \right)$ for $j=1,2$, $n=-1,0,...,k$ are numbers which can be determined from tables or by calculators. Next, we use algorithm \ref{alg:genhhl} for solving equation \ref{mateq} with entries given in formula \ref{mentries} and refer reader to \cite{SarFirst} for more details on numerical experiment in Qiskit.

\subsection{HHL algorithm for approximate solution of the Inverse Two-Phase Stefan Problem}
\label{sec:approx}
In previous section we considered $\alpha\sqrt{t}$ case, as for arbitrary $\alpha(t)$ boundary we follow same principle and use Fa Di Bruno's Formula in combination with HHL quantum algorithm to find exact solutions. Worth noting that, in electrical engineering, it's sometimes sufficient and useful to utilize approximate solutions of the problems where error can be estimated using the Maximum principle.\\ 
In this section we will demonstrate the use of HHL algorithm for approximate solution (collocation method) of the Inverse Two-Phase Stefan Problem which is used for modeling arcing processes and determining heat flux function \cite{sar17filomat}. Direct Stefan problem is a type of free boundary value problems where along with a temperature function $\theta$ in \ref{mod1_eq1} and \ref{mod1_eq2}, an unknown moving boundary has to be determined. In inverse Stefan problems moving boundary is given and known, the goal is to reconstruct functions at boundary conditions and temperature functions in system of Heat Equations. These problems are widely used for modeling wide range of transient phenomena in chemistry, physics, biology and economics \cite{Tarzia2000ABO,Friedman00freeboundary}. In the problem below, moving boundary $\alpha(t)$ is given and besides temperature function $\theta$ in \ref{mod2_eq1},\ref{mod2_eq2}, flux function $P(t)$ has to be determined. Let's consider following linear Inverse Two-Phase Stefan Problem
\begin{align}
&\frac{\partial {\theta_{1}}}{\partial {t}} = a_{1}^{2}\frac{\partial {\theta_{1}^{2}}}{\partial {x^{2}}},\hspace{1em}  &0<x<\alpha(t),\hspace{1em} 0<t<t_{a} \label{mod2_eq1}\\[1mm]
&\frac{\partial {\theta_{2}}}{\partial {t}} = a_{2}^{2}\frac{\partial {\theta_{2}^{2}}}{\partial {x^2}},\hspace{1em}  &\alpha(t)<x<X,\hspace{1em} 0<t<t_{a} \label{mod2_eq2}\\[3mm]
&\theta_{1}(0,0) = T_{m},\label{mod2_incond1}\\[3mm]
&\theta_{2}(x,0) = f(x),\label{mod2_incond2}\\[3mm]
&f(0)=T_{m},\hspace{1mm}\alpha(0)=0,\hspace{1mm}\underset{x\to \infty}{\lim}f(x)\approx f(X)=0,\hspace{1mm} \hspace{1mm}&\underset{x\to \infty}{\lim}\theta(x,t)\approx \theta(X,t)=0,\label{concord}\\[3mm]
&\left.-\lambda_{1}\frac{\partial{\theta_1}}{\partial{x}} \right|_{x=0} = P(t),\label{mod2_flux}\\[3mm]
&\theta_{1}(\alpha(t),t) = \theta_{2}(\alpha(t),t)=T_{m}  \label{mod2_temp}\\[3mm]
&\left.-\lambda_{1}\frac{\partial \theta_{1}}{\partial x} \right|_{x=\alpha(t)} = \left.-\lambda_{2}\frac{\partial \theta_{2}}{\partial x} \right|_{x=\alpha(t)}+L\gamma\frac{\partial \alpha(t)}{\partial t}\label{mod2_stef}
\end{align}
where $T_m$, $X$ and $t_{a}$ are melting temperature, finite electric contact radius and arcing duration respectively.\\
Following analogy in section \ref{sec:exact} we represent solution in the form of series
\begin{align}
&U_1(x,t)=T_m + \sum_{n=0}^{k}\left(2a_{1}\sqrt{t}\right)^{n}\left[A_ni^nerfc\left( \frac{x}{2a_{1}\sqrt{t}} \right)+ B_ni^nerfc\left( \frac{-x}{2a_{1}\sqrt{t}} \right)\right], \label{mod2_sol1}\\[3mm]
&U_2(x,t)= T_m + \sum_{n=0}^{k}\left(2a_{2}\sqrt{t}\right)^{n}\left[C_ni^nerfc\left( \frac{x}{2a_{2}\sqrt{t}} \right)+ D_ni^nerfc\left( \frac{-x}{2a_{2}\sqrt{t}} \right)\right], \label{mod2_sol2}
\end{align}
where coefficients $D_{n}$ at \ref{mod2_sol2} can be found from \ref{mod2_incond2} following the same principle in section \ref{sec:approx}.\\Thus,
\begin{align}
&D_n=\frac{f^{n}(0)}{2}.
\label{mod2_Dn}
\end{align}
Let $P(t)=\sum_{n=0}^{k}P_{n}t^n$, where $P_n=\frac{P^{n}(0)}{n!}$ and have to be determined from boundary and initial conditions. The idea of the collocation method applied in this problem is to subdivide $0<t<T_a$ into $k$ intervals and after substituting solution functions \ref{mod2_sol1}, \ref{mod2_sol2} into the boundary conditions \ref{mod2_flux},\ref{mod2_temp},\ref{mod2_stef} at points $t_1, t_2,...,t_k$ solve the system of linear algebraic equation or matrix equation \ref{mateq} for coefficients $A_n, B_n, C_n, P_n$ using HHL algorithm, where $M, x, b$ in equation \ref{mateq} are as following:
\begin{align}
&M=\begin{pmatrix}m_{1,1}&\dots&m_{1,4}&\dots&m_{{1,}{4k-3}}&\dots&m_{{1,}{4k}}\\\vdots\\m_{k,1}&\dots&m_{k,4}&\dots&m_{{k,}{4k-3}}&\dots&m_{{k,}{4k}}\\m_{k+1,1}&\dots&m_{k+1,4}&\dots&m_{{k+1,}{4k-3}}&\dots&m_{{k+1,}{4k}}\\\vdots\\m_{2k,1}&\dots&m_{2k,4}&\dots&m_{{2k,}{4k-3}}&\dots&m_{{2k,}{4k}}\\m_{2k+1,1}&\dots&m_{2k+1,4}&\dots&m_{{2k+1,}{4k-3}}&\dots&m_{{2k+1,}{4k}}\\\vdots\\m_{3k,1}&\dots&m_{3k,4}&\dots&m_{{3k,}{4k-3}}&\dots&m_{{3k,}{4k}}\\m_{3k+1,1}&\dots&m_{3k+1,4}&\dots&m_{{3k+1,}{4k-3}}&\dots&m_{{3k+1,}{4k}}\\\vdots\\m_{4k,1}&\dots&m_{4k,4}&\dots&m_{{4k,}{4k-3}}&\dots&m_{{4k,}{4k}}\end{pmatrix},\label{stefmentries}
\end{align}
\begin{align*}
\\&x=\begin{pmatrix}A_0\\B_0\\C_0\\P_0\\\vdots\\A_k\\B_k\\C_k\\P_k\end{pmatrix},
b=\begin{pmatrix}0\\\vdots\\0\\0\\\vdots\\0\\-\left(2a_{2}\sqrt{t_{1}}\right)^{k}i^{k}erfc\left( \frac{-\alpha(t_1)}{\left(2a_{2}\sqrt{t_{1}}\right)} \right)\\\vdots\\-\left(2a_{2}\sqrt{t_{k}}\right)^{k}i^{k}erfc\left( \frac{-\alpha(t_k)}{\left(2a_{2}\sqrt{t_{k}}\right)} \right)\\L\gamma\left.\frac{d\alpha(t)}{dt}\right|_{t_{1}}-\lambda_2\frac{\left(2a_{2}\sqrt{t_{1}}\right)^{k-1}}{\left(\alpha(t_1)\right)^2}i^{k-1}erfc\left( \frac{\alpha(t_1)}{2a_{2}\sqrt{t_{1}}} \right)\\\vdots\\L\gamma\left.\frac{d\alpha(t)}{dt}\right|_{t_{k}}-\lambda_2\frac{\left(2a_{2}\sqrt{t_{k}}\right)^{k-1}}{\left(\alpha(t_k)\right)^2}i^{k-1}erfc\left( \frac{\alpha(t_k)}{2a_{2}\sqrt{t_{k}}} \right)\end{pmatrix},
\end{align*}
where
\begin{align*}
&m_{k,4k-3}=-\lambda_1 \frac{\left(2a_{1}\sqrt{t_{k}}\right)^{k-1}}{\alpha(t_k)}i^{k-1}erfc\left( 0 \right),\\ 
&m_{k,4k-2}=-\lambda_1 \frac{\left(2a_{1}\sqrt{t_{k}}\right)^{k-1}}{\alpha(t_k)}i^{k-1}erfc\left( 0 \right),\\
&m_{k,4k-1}=0,
m_{k,4k}=t_k^(k-1),\\
&m_{2k,4k-3}=\frac{\left(2a_{1}\sqrt{t_{k}}\right)^{k}}{\alpha(t_k)}i^{k}erfc\left( \frac{\alpha(t_k)}{2a_{1}\sqrt{t_{k}}} \right),\\
&m_{2k,4k-2}=\frac{\left(2a_{1}\sqrt{t_{k}}\right)^{k}}{\alpha(t_k)}i^{k}erfc\left( \frac{-\alpha(t_k)}{2a_{1}\sqrt{t_{k}}} \right),\\
&m_{2k,4k-1}=0,m_{2k,4k}=0,\\
&m_{3k,4k-3}= 0,m_{3k,4k-2}=0,\\
&m_{3k,4k-1}= \frac{\left(2a_{2}\sqrt{t_{k}}\right)^{k}}{\alpha(t_k)}i^{k}erfc\left( \frac{\alpha(t_k)}{2a_{2}\sqrt{t_{k}}} \right),\\
&m_{3k,4k}= 0,\\
&m_{4k,4k-3}=\lambda_1\frac{\left(2a_{1}\sqrt{t_{k}}\right)^{k-1}}{\alpha(t_k)}i^{k-1}erfc\left( \frac{\alpha(t_k)}{2a_{1}\sqrt{t_{k}}} \right),\\ 
&m_{4k,4k-2}=-\lambda_1\frac{\left(2a_{1}\sqrt{t_{k}}\right)^{k-1}}{\alpha(t_k)}i^{k-1}erfc\left( -\frac{\alpha(t_k)}{2a_{1}\sqrt{t_{k}}} \right),\\
&m_{4k,4k-1}=-\lambda_2\frac{\left(2a_{2}\sqrt{t_{k}}\right)^{k-1}}{\left(\alpha(t_k)\right)^2}i^{k-1}erfc\left( \frac{\alpha(t_k)}{2a_{2}\sqrt{t_{k}}} \right),\\
&m_{4k,4k}=0,\\
\end{align*}
and expressions $\pm\lambda_j\frac{\left(2a_{j}\sqrt{t_{n}}\right)^{n-1}}{\alpha(t_n)}i^{n-1}erfc\left( \frac{-\alpha(t_n)}{2a_{2}\sqrt{t_{n}}} \right)$, 
$\pm\lambda_j\frac{\left(2a_{j}\sqrt{t_{n}}\right)^{n-1}}{\alpha(t_n)}i^{n-1}erfc\left( 0 \right)$ \\
where $j=1,2$, $n=0,...,k$ are numbers which can be determined from tables.  Next, we use Quantum HHL algorithm \ref{alg:genhhl} for solving problem \ref{mateq} with entries given in \ref{stefmentries}. Numerical implementation is demonstrated in \cite{SarFirst}.





\section{Experimental Results and Discussion} 
We used IBM Q and Qiskit for experiments and programming purposes. MBVP and the Inverse Two-Phase Stefan Problem were solved with fidelities 0.99 and 1 respectively. We refer reader to \cite{SarFirst} for details of experiments. 
Proposed method in combination with Fa Di Bruno's Formula and Quantum HHL algorithm can be used for exact solutions for direct/inverse Stefan type problems and MBVPs in general for arbitrary $\nu$ in \ref{geneq} and arbitrary $\alpha(t)$.
Special functions method in combination with HHL algorithm or its Continuous Variable version \cite{Arrazola_2019} can be also used for approximate solutions of boundary value problems with fixed boundaries as well.       

\section{Conclusions}
\label{sec:conclusions}
HHL quantum algorithm was used for exact and approximate solutions of moving boundary value problems. We used IBM Q for experiments \cite{SarFirst} and solved MBVP with discontinuous coefficients and Inverse Two-Phase Stefan problem. $A_0,A_1, B_0,B_1,C_0, C_1$ and $A_0,A_1, B_0,B_1,C_0, C_1,P_0,P_1$ coefficients of solution functions in \ref{mod1_sol1},\ref{mod1_sol2} and \ref{mod2_sol1},\ref{mod2_sol2} were found with fidelities 0.99 and 1 respectively.




\bibliography{mybibfile}

\end{document}